\documentclass[12pt,a4paper]{article}
\usepackage{authblk}
\usepackage[top=2cm, bottom=2cm, left=2cm, right=2cm]{geometry}
\usepackage{fancyhdr}
\usepackage{cite}
\usepackage{amsmath,amssymb,amsfonts,amsthm,makeidx,graphicx}
\usepackage{txfonts}
\usepackage{helvet}
\usepackage[]{mdframed}
\usepackage{url}

\renewenvironment{abstract}{%
\hfill\begin{minipage}{0.95\textwidth}
\rule{\textwidth}{1pt}}
{\par\noindent\rule{\textwidth}{1pt}\end{minipage}}
\makeatletter
\renewcommand\@maketitle{%
\hfill
\begin{minipage}{0.95\textwidth}
\vskip 2em
\let\footnote\thanks 
{\LARGE \@title \par }
\vskip 1.5em
{\large \@author \par}
\end{minipage}
\vskip 1em \par
}
\makeatother
\begin{document}
%
\title{Submodular Maximization Subject to Uniform and Partition Matroids: From Theory to Practical Applications and Distributed Solutions}
\author[1]{Solmaz S. Kia}
\affil[1]{{\small University of California Irvine, Mechanical and Aerospace Engineering Dept., 4200 Engineering Gateway, Irvine, CA 92697, USA. email: solmaz@uci.edu}}

\newtheorem{thm}{Theorem}[section]
\newtheorem{prop}{Proposition}[section]
\newtheorem{rem}{Remark}[section]
\newtheorem{cor}{Corollary}[section]
\newtheorem{lem}{Lemma}[section]
\newtheorem{definition}{Definition}
\maketitle
\begin{abstract}
This article provides a comprehensive exploration of submodular maximization problems, focusing on those subject to uniform and partition matroids. Crucial for a wide array of applications in fields ranging from computer science to systems engineering, submodular maximization entails selecting elements from a discrete set to optimize a submodular utility function under certain constraints. We explore the foundational aspects of submodular functions and matroids, outlining their core properties and illustrating their application through various optimization scenarios. Central to our exposition is the discussion on algorithmic strategies, particularly the sequential greedy algorithm and its efficacy under matroid constraints. Additionally, we extend our analysis to distributed submodular maximization, highlighting the challenges and solutions for large-scale, distributed optimization problems. This work aims to succinctly bridge the gap between theoretical insights and practical applications in submodular maximization, providing a solid foundation for researchers navigating this intricate domain.
\end{abstract}

\bigskip
\begin{minipage}{0.1\textwidth}
\end{minipage}
\fbox{\begin{minipage}{0.95\textwidth}
-- Submodular Functions and Practical Applications: The article highlights how submodular maximization, especially with matroid constraints, serves as a cornerstone for solving various optimization problems in computer science, systems engineering, and beyond. Its applications range from data summarization and sensor placement to network resource management, demonstrating submodularity's wide-ranging impact.

~\\
-- Algorithmic Strategies for Optimization: A significant portion of the article explores crucial algorithmic approaches like the sequential greedy algorithm and the continuous greedy algorithm. These methods are not only foundational for achieving polynomial-time solutions to submodular maximization problems but also exhibit guaranteed performance bounds, making them vital tools in optimization theory.

~\\
-- Adaptations for Distributed Environments: The discussion extends to the adaptation of submodular maximization strategies in distributed settings, reflecting the growing need for optimization solutions that cater to privacy concerns and the decentralized nature of modern computational and network systems. 

~\\
-- Innovative Research Directions: Lastly, the article concludes with an overview of emerging research areas within the field, such as deep submodular functions, online maximization, and fairness in optimization. These forefront research topics signal the ongoing evolution of submodular maximization theory and its expanding applicability to new, complex problems.
\end{minipage}}

\section{Introduction}\label{chap1:sec1}

In the modern landscape of intelligent systems, the demand for optimization algorithms that facilitate optimal decision-making for efficient resource allocation and policy making is at an all-time high. Optimization algorithms enabling these optimal decision-makings are expected to be computationally efficient and come with reasonable communication cost to deliver timely and efficient solutions for in-network operations. Some of the optimal resource allocation and policy-making problems faced for such systems are in the form of a combinatorial optimization problem
\begin{equation}\label{eq::gen_set_opt}
    \max f(\mathcal{S})~~\text{subject to~}\mathcal{S}\in\mathcal{F}(\mathcal{P}),
\end{equation}
where the goal is to choose a subset $\mathcal{S}$ of discrete elements from a ground set $\mathcal{P}$ that maximizes a utility, described by function $f:2^\mathcal{P}\to\mathbb{R}_{\geq0}$  while respecting constraints captured by $\mathcal{F}$, the discrete set of feasible solutions. 

\medskip
Combinatorial optimization problems of the form~\eqref{eq::gen_set_opt} are often NP-hard, meaning that their complexity class are intrinsically harder than those that can be solved by a nondeterministic Turing machine in polynomial time. The quest thus is focused on finding suboptimal solutions with guaranteed optimality gap that measures the distance between the suboptimal solution and the optimal one. That is, we seek polynomial time suboptimal solutions with well-characterized optimality gap $0< \alpha <1$ satisfying 
\begin{equation}\label{eq::optimality_gap}
   \alpha f(\mathcal{S}^\star)\leq f(\bar{\mathcal{S}})\leq  f(\mathcal{S}^\star),
\end{equation}
where $\mathcal{S}^\star$ is the global maximizer of~\eqref{eq::gen_set_opt} and $\bar{\mathcal{S}}$ is the decision set delivered by the suboptimal solver. 

For a certain class of optimization problem~\eqref{eq::gen_set_opt}, namely problems with monotone submodular utility functions subject to matroid constraints, there is a significant body of work in the literature that provides suboptimal solutions with fully characterized optimality gap when the problem is solved in a centralized manner. In a distributed setting, either the utility function $f$ or the elements of the ground set $\mathcal{P}$ in~\eqref{eq::gen_set_opt}  are fragmented and distributed among the players/agents (devices/subsystems) of the optimization problem. Distributed operations bring about new set of challenges that can exasperate the established optimality gaps and require special considerations to manage the inter-agent communication costs. Additional concerns, such as scalability and the demand for privacy preservation, compound the complexity of distributed algorithms. Distributed algorithms for submodular maximization are not as extensively explored and established as the centralized algorithms.

\medskip
Submodular maximization theory has been widely acknowledged as an effective approach for solving combinatorial resource allocation problems in the field of computer science. Many problems such as agglomerative clustering, exemplar-based clustering~\cite{PH-AS-SB-KM:21}, categorical feature compression~\cite{AR-HE-LC-MB-TF-VM:19}, 
recommender systems~\cite{KECG:11}, search result diversification~\cite{AB-AJ-HCL-YY:17
}, data subset selection~\cite{KW-RI-JB:15},
social networks analysis~\cite{QN-JG-CH-WW:20} are cast as submodular maximization problems subject to matroid constraints. Classical combinatorial optimization problems like minimum spanning tree, global minimum cut, maximum matching, traveling salesman problem, max clique, max cut, set cover and knapsack, among the others, are also another set of combinatorial optimization problems that can be modeled as submodular maximization problems. These classical problems often appear as optimal decision-making mechanism in many system and engineering problems. Thus, robust solution to the NP-hard submodular maximization problems can serve as the backbone of optimal decision-making in modern industries such as transportation, supply chain, energy, finance, and scheduling. However, submodular maximization problems may not be as well known in the systems and control field as in computer science field. Submodular maximization generalizes many classic problems in combinatorial optimization and has recently found a wide range of applications in operational planning/task. Some example problems include sensor and actuator placement problems~\cite{AK-AS-CG:08,THS-FLC-JL:16,ZL-AC-PL-LB-DK-RP:18,AC-PL-BA-LB-RP:18,AK-CG:07,AC-LB-RP:14}, energy storage placement~\cite{JQ-IY-RR:19,MB-SP-ED-AV:20}, measurement scheduling~\cite{STJ-SLS:15}, voltage control in smart grid~\cite{ZL-AC-PL-LB-DK-RP:16}, persistent monitoring via mobile robots~\cite{NR-SSK:21}. 

This article will focus on solution approaches, application examples and distributed implementation of the two most well-know submodular maximization problems that are selecting elements from a finite discrete ground set $\mathcal{P}$ to maximize a submodular utility function, namely 
\begin{itemize}
    \item \emph{Submodular maximization subject to uniform matroid constraint}
\begin{equation}\label{eq::submdoular_uniform}
    \mathcal{S}^\star =\arg\max_{\mathcal{S}\subset\mathcal{P}} f(\mathcal{S})~~\text{subject to~} |\mathcal{S}|\leq \kappa,
\end{equation}
for a given $\kappa\in\mathbb{Z}_{\geq1}$.
    \item \emph{Submodular maximization subject to partition matroid constraint}
    \begin{equation}\label{eq::submdoular_partition}
    \mathcal{S}^\star =\arg\max_{\mathcal{S}\subset\mathcal{P}} f(\mathcal{S})~~\text{subject to~} |\mathcal{S}\cap \mathcal{P}_i|\leq \kappa_i, ~i\in\{1,\cdots,N\}
\end{equation}
or a given $\kappa\in\mathbb{Z}_{\geq1}$. Here,  $\mathcal{P}\cup_{i=1}^N\mathcal{P}_i$ and $\mathcal{P}_i\cap\mathcal{P}_j=\emptyset$ for all $i\neq j$.
\end{itemize}
\medskip
The remainder of this article is organized as follows. 
Section~\ref{eq:submodular_functions} provides a brief overview of the fundamental properties of the submodular functions and matroid constraints. Section~\ref{sec::applications}  provides some example applications of submodular maximization theory. Section~\ref{sec:central_solutions} reviews the existing centralized optimization algorithms to solve submodular maximization problems. Section~\ref{sec::distributed} reviews various distributed submodular maximization problems, the challenges in solving these problems and some existing algorithms and their design methods. And finally, Section~\ref{se:conclusion} gives our concluding remarks.

\section{Submodular functions and matroid constraints}\label{eq:submodular_functions}

\begin{figure}[t]
\begin{center}
    {\includegraphics[width=.7\textwidth]{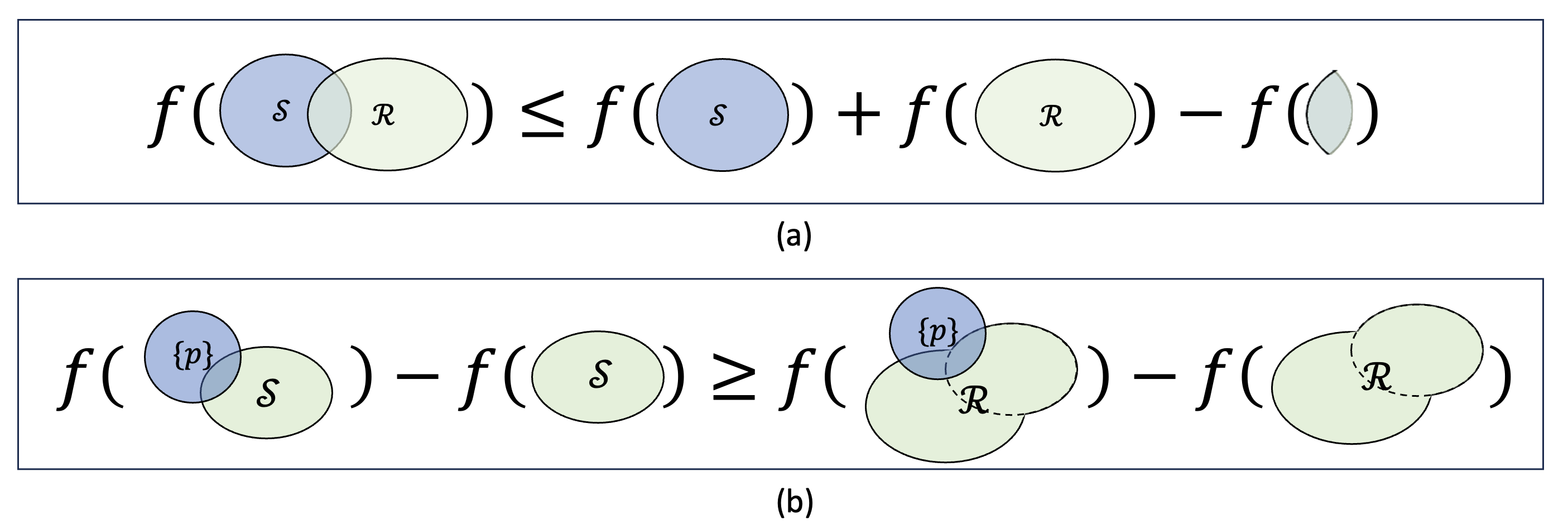}}%
    \caption{{\small   
    Properties of submodular functions.
}}\label{fig:city_partition}
\end{center}
\end{figure}

In what follows, we introduce what submodular function and matroid constraints are. We provide an overview of the relevant components of these concepts pertaining to the two submodular maximization problems~\eqref{eq::submdoular_uniform} and \eqref{eq::submdoular_partition}, which we want to discuss in this article. Interested readers can consult references such as~\cite{CHP=KS:98} and~\cite{SF:05} for a more detailed and comprehensive overview. 

Consider a finite ground set $\mathcal{P}$ of $n$ discrete elements, which without loss of generality, is assumed to be $\mathcal{P}=\{1,\cdots,n\}$. 

\begin{definition}[Normal set function]
    A set function $f: 2^\mathcal{P}\to\mathbb{R}_{\geq 0}$ is said to be normal if $f(\emptyset)=0$.
\end{definition}
\begin{definition}[Monotone increasing set function]
    A set function $f: 2^\mathcal{P}\to\mathbb{R}_{\geq 0}$ is said monotone increasing if and only if 
\begin{align*}
    f(\mathcal{S})\geq f(\mathcal{R}),\quad \forall \mathcal{R}\subset\mathcal{S}\subset\mathcal{P}.
\end{align*}
\end{definition}
\begin{definition}[Submodular function]
    The function $f: 2^\mathcal{P}\to\mathbb{R}_{\geq0}$ is submodular if and only if
\begin{align}\label{eq::submodular_main2}
    f(\mathcal{R})+f(\mathcal{S}) \geq  f(\mathcal{R}\cup\mathcal{S})+f(\mathcal{R}\cap \mathcal{S}),\quad \forall \,\mathcal{S},\mathcal{R}\in\mathcal{P}.
\end{align}
\end{definition}
The relationship~\eqref{eq::submodular_main2} fundamentally highlights how submodular functions evaluate redundancy or overlap between sets. Specifically,~\eqref{eq::submodular_main2} suggests that if a utility function is submodular, the loss of utility at joint evaluation of any two subsets, i.e., at $\mathcal{R}\cup \mathcal{S}$, of the ground set  compared to the sum of the utility of the individual subsets $\mathcal{R}$ and $\mathcal{S}$ is more than or equal to the utility of the overlap $\mathcal{R}\cap \mathcal{S}$, see Fig.~\ref{fig:city_partition}(a). Thus, the relation elegantly penalizes redundancy or overlap within the sets' arguments, quantitatively expressing that the function $f$ attributes a lower incremental value to shared or overlapping elements between $\mathcal{R}$ and $\mathcal{S}$, effectively encouraging diversity within the sets. When this loss is equal to the utility of the overlap the set function $f$ is said to be \emph{modular}.
\begin{definition}[Modular set function]
    The function $f: 2^\mathcal{P}\to\mathbb{R}_{\geq0}$ is submodular if and only if
    \begin{align}\label{eq::submodular_main}
    f(\mathcal{R})+f(\mathcal{S}) =  f(\mathcal{R}\cup\mathcal{S})+f(\mathcal{R}\cap \mathcal{S}),\quad \forall \,\mathcal{S},\mathcal{R}\in\mathcal{P}.
\end{align}
\end{definition} 
 If $f$ is modular then there is a weight function $w : \mathcal{P} \to\mathbb{R}_{\geq0}$ such that $f(\mathcal{R}) =  \sum_{r\in\mathcal{R}}w(r)$. 

For a set value function $f:2^\mathcal{P}\to\mathbb{R}_{\geq0}$, the \emph{marginal gain}, 
\begin{align}
   \Delta_f(p|{\mathcal{S}})= f(\mathcal{S}\cup\{p\})-f(\mathcal{S}),
\end{align} quantifies how much the function $f$, which maps sets to real numbers, increases as a result of adding an element $p\in\mathcal{P}$ to $\mathcal{S}\subset\mathcal{P}$. Marginal gain often is also called the \emph{discrete derivative} of $f$ at $\mathcal{S}$ with respect to $p$. In the context of $f$ being a utility function, marginal gain measures the additional value or utility obtained by including the element $p$ into the set $\mathcal{S}$. Marginal gain is a crucial concept in optimization, especially when dealing with set functions, because it helps in understanding the benefit of including an additional element in a set given the current context, i.e., the elements already selected in the set $\mathcal{S}$. Submodular functions are  characterized by their diminishing return, i.e., a set function $f:2^\mathcal{P}\to\mathbb{R}_{\geq0}$ is submodular if and only if 
\begin{align}\label{eq::submodular_marginal}
    f(\mathcal{S}\cup \{p\})-f(\mathcal{S})\geq f(\mathcal{R}\cup \{p\})-f(\mathcal{R}), \quad\forall \,\mathcal{S}\subset\mathcal{R}\subset\mathcal{P},~p\in\mathcal{P}\backslash\mathcal{R};
\end{align}
see Fig.~\ref{fig:city_partition}(b). 
The diminishing return property makes submodular functions particularly appealing for problems where resources are being allocated, tasks are being assigned, or selections are being made from a set, and the goal is to maximize some notion of utility or coverage subject to constraints (like size, budget, or time). This property ensures that as solutions grow, the incremental gains diminish, leading to natural ``saturation" points that help in defining ``optimal" or ``near-optimal" solutions to such problems. For the special case of modular functions, the utility of adding an element to a set is independent of the set's current composition—paralleling linear functions in numerical optimization. That is for modular functions
\begin{align}\label{eq::modular_marginal}
    f(\mathcal{S}\cup \{p\})-f(\mathcal{S})=f(\mathcal{R}\cup \{p\})-f(\mathcal{R}), \quad\forall \,\mathcal{S}\subset\mathcal{R}\subset\mathcal{P},~p\in\mathcal{P}\backslash\mathcal{R};
\end{align}

For any monotone increasing and submodular function $f:2^\mathcal{P}\to\mathbb{R}_{\geq0}$, the total \emph{curvature} $c\in[0,1]$ of the function, defined as 
\begin{align}\label{eq::totalCurvature}
    c = 1 - \underset{\mathcal{R} \subset \mathcal{P},\,\,p \not \in \mathcal{R}}{\textup{min}}\frac{\Delta_f(p|{\mathcal{R}})}{\Delta_f(p|\emptyset)}.
\end{align}
This definition aims to capture the least proportional increase in the function value when adding an element to any set, relative to adding it to the empty set. For modular functions,\eqref{eq::modular_marginal} implied that $c=0$. Thus, the total curvature $c$ of a submodular function is a measure of how far the function is from being linear. On the other hand, if $c = 1$, this typically indicates the case where the function experiences the greatest reduction in marginal gains for adding elements to a set. Specifically, it signifies that there is at least one situation where adding an element to a larger set grants no additional value compared to adding it to some smaller set, thus reflecting a scenario with the maximum possible loss in marginal gain as the set size increases. 

For a normal, monotone increasing and submodular function $f$, we can derive a tighter bound on the marginal gain. Specifically, for any $\mathcal{S}\subset\mathcal{P}$ and any $r\in\mathcal{P}$, from the definition~\eqref{eq::totalCurvature} of the total curvature, we can write
\begin{align*}
&  \frac{f(\mathcal{S}\cup r)-f(\mathcal{S})}{f(r)}\geq \min_{\mathcal{S}\subset\mathcal{P},p\notin\mathcal{S})}\frac{f(\mathcal{S}\cup p)-f(\mathcal{S})}{f(p)}=1-c~\rightarrow ~f(\mathcal{S}\cup r)-f(\mathcal{S})\geq (1-c)f(r).
\end{align*}
Building on this result, we can extend our analysis to consider the addition of multiple elements. For any $\mathcal{S}\subset\mathcal{P}$, and $\mathcal{R}=\{r_1,\cdots,r_m\}\subset\mathcal{P}$, we can derive the following inequalities
 \begin{align*}
       & f(\mathcal{S} \cup\{r_1,\dots,r_m\})- f(\mathcal{S} \cup\{r_1,\dots,r_{m-1}\})\geq (1-c)f(r_m)\\
     &   f(\mathcal{S} \cup\{r_1,\dots,r_{m-1}\})- f(\mathcal{S} \cup\{r_1,\dots,r_{m-2})\geq (1-c)f(r_{m-1})\\
     &\vdots\\
    &  f(\mathcal{S} \cup\{r_1\})- f(\mathcal{S})\geq (1-c)f(r_{1}).
    \end{align*}
By summing these inequalities and applying the submodularity of $f$, we arrive at a more general result
    \begin{equation}
        f(\mathcal{S}\cup\mathcal{R})-f(\mathcal{S})\geq (1-c)\sum\nolimits_{i=1}^mf(r_i)\geq (1-c)f(\mathcal{R}).
    \end{equation}
This inequality provides a tighter bound on the marginal gain when adding a set $\mathcal{R}$ to $\mathcal{S}$, taking into account the total curvature of the function. As a consequence, for any normal monotone increasing and submodular function, we can conclude that
\begin{equation}\label{eq::curvture_inequality2}
    f(\mathcal{R}\cup\mathcal{S})-f(\mathcal{R})\geq f(\mathcal{S})-cf(\mathcal{R}).
    \end{equation}

\begin{figure}[t]
\begin{center}
    {\includegraphics[width=.5\textwidth]{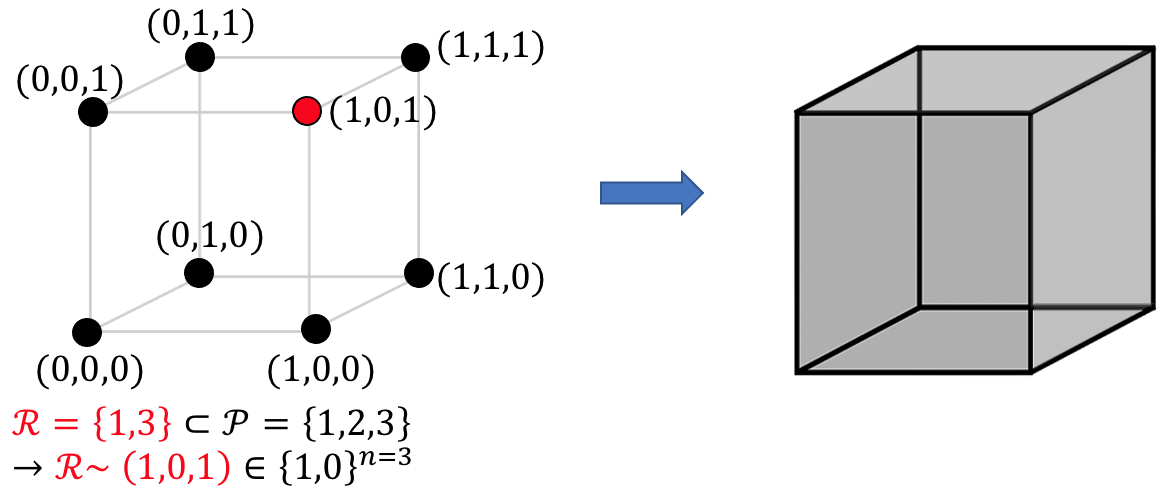}}%
    \caption{{\small   
    Multilinear extension extends a submodular function $f(\mathcal{R})$ to the continuous space defined on hypercube $[0,1]^n$. $F(\boldsymbol{x})$ agrees with $f(\mathcal{R})$ on the vertices of the hypercube (for integral $\boldsymbol{x}$).
}}\label{fig:multi_linear}
\end{center}
\end{figure}

\medskip
Having explored the implications of total curvature on submodular functions, we now turn our attention to other important properties of submodularity. These properties are fundamental in understanding how submodular functions behave under various operations, which is crucial for their application in diverse optimization problems. Firstly, submodularity is preserved under taking nonnegative linear combinations, i.e., if $f_1,\cdots,f_m: 2^\mathcal{P}\to\mathbb{R}_{\geq0}$ is submodular then $f(\mathcal{S})=\sum_{i=1}^m \alpha_i f_i(\mathcal{S})$ is submodular for any $\alpha_1,\cdots,\alpha_m\in\mathbb{R}_{\geq0}$. Secondly, monotone submodularity is preserved under truncation, i.e., if $f: 2^\mathcal{P}\to\mathbb{R}_{\geq0}$ is submodular, then so is $f(\mathcal{S})=\min\{f(\mathcal{S}),c\}$ for any constant $c\in\mathbb{R}$. Lastly, submodularity is preserved when we take the residual, i.e.,  if $f: 2^\mathcal{P}\to\mathbb{R}_{\geq0}$ is submodular, and $\mathcal{R},\mathcal{S}\subset\mathcal{P}$ are any disjoint sets, then the residual $g : 2^\mathcal{R}\to\mathbb{R}_{\geq0}$ defined as $g(\mathcal{Q}) = f(\mathcal{S}\cup\mathcal{Q} ) -f(\mathcal{Q})$ for any $\mathcal{Q}\subset\mathcal{R}$ is submodular.

\medskip
Set functions with ground set of $n$ elements can be cast as a vector function defined on vertices of hypercube $\{0,1\}^n$ as shown in Fig.~\ref{fig:multi_linear}.~\cite{JV:08} proposed to use the \emph{multilinear extension} of submodular set functions to extend the submodular utility function $f:2^{\mathcal{P}} \to \mathbb{R}_{\geq0}$, which is defined on the vertices of the $n-$dimensional hypercube $\{0,1\}^n$, to the continuous multilinear function $F(\boldsymbol{x})$ defined as 
\begin{align}\label{eq::F_determin}
    F(\boldsymbol{x}) = \sum_{\mathcal{R} \subset \mathcal{P}}{} f(\mathcal{R}) \prod_{p \in \mathcal{R}}^{} [\boldsymbol{x}]_p \prod_{p \not\in \mathcal{R}}^{}(1-[\boldsymbol{x}]_p),~~ \boldsymbol{x} \in [0,1]^{n},
\end{align}
in the continuous space, where $[\boldsymbol{x}]_p$ is the $p$th element of $\boldsymbol{x}$. The multilinear extension agrees with $f(\mathcal{S})$ ar the integral values of $\boldsymbol{x}$. More precisely, for any $\mathcal{R}\subset\mathcal{P}$, we have  $F(\boldsymbol{1}_\mathcal{R})=f(\mathcal{R})$ where $\boldsymbol{1}_\mathcal{R}$ is a vector whose elements corresponding to set $\mathcal{R}$ are $1$ and the rest are $0$. This multilinear extension plays a crucial role in solving submodular maximization problems through a technique known as continuous relaxation, which will be elaborated upon in Section~\eqref{sec:central_solutions}.

\medskip
Multilinear extension essentially allows each element's inclusion in a set to be considered as a probability, smoothing the transition from discrete to continuous space for optimization. More precisely, interpreting  $\boldsymbol{x} \in[0,1]^n$ as \emph{the membership probability vector} and positing $\mathcal{R}_{\boldsymbol{x}}\subset\mathcal{P}$ as a random set where each element $p \in\mathcal{P}=\{1,\cdots,n\}$ is included in $\mathcal{R}_{\boldsymbol{x}}$ with probability $[\boldsymbol{x}]_p$, equation \eqref{eq::F_determin} translates to
\begin{align}\label{eq::F_stoc}
   F(\boldsymbol{x})=\mathbb{E}[f(\mathcal{R}_{\boldsymbol{x}})].  
\end{align} 
Vondrák's work reveals that the first derivative of $F(\boldsymbol{x})$ with respect to $[\boldsymbol{x}]_p$, as shown in
\begin{align}\label{eq::firstDer}
\frac{\partial F}{\partial [\boldsymbol{x}]_p} (\boldsymbol{x})= \mathbb{E}[f(\mathcal{R}_{\boldsymbol{x}} \cup \{p\})-f(\mathcal{R}_{\boldsymbol{x}} \setminus \{p\})],
\end{align}
consistently remains non-negative for monotone increasing submodular functions. Furthermore, the analysis extends to the second derivative of $F(\boldsymbol{x})$, indicating
\begin{align}\label{eq::secondDer}
& \frac{\partial^2 F}{\partial [\boldsymbol{x}]_p \partial [\boldsymbol{x}]_q}(\boldsymbol{x}) = \mathbb{E}[f(\mathcal{R}_{\boldsymbol{x}} \cup \{p,q\})-f(\mathcal{R}_{\boldsymbol{x}} \cup \{q\} \setminus \{p\})-f(\mathcal{R}_{\boldsymbol{x}} \cup \{p\} \setminus \{q\})+f(\mathcal{R}_{\boldsymbol{x}} \setminus \{p,q\})],
\end{align}
which, by virtue of submodularity, guarantees $ \frac{\partial^2 F}{\partial [\boldsymbol{x}]_p \partial [\boldsymbol{x}]_q}(\boldsymbol{x})\leq 0$. 

\smallskip
The multilinear extension function, as detailed in equation~\eqref{eq::F_determin}, is positioned within the broader category of continuous Diminishing Returns (DR) submodular functions. These functions constitute a distinct subclass within the spectrum of non-convex/non-concave continuous functions. The problem of optimizing DR-submodular functions over a convex set has attracted considerable interest in both the machine learning and theoretical computer science communities due to its wide applicability in data mining, machine learning, economics, and statistics. Readers keen on a deeper exploration of this domain may find valuable insights in seminal work such as~\cite{bian2017continuous} or recent work such as~\cite{MP-CQ-VA:24}.

\medskip

Having reviewed the submodular functions, we next introduce \emph{matroids}, which play a crucial role in addressing complex problems such as submodular maximization, particularly when constrained by a uniform matroid, a fundamental concept in combinatorial optimization. Matroids provide a powerful abstract framework that generalizes the notion of linear independence from vector spaces to more versatile combinatorial structures.
A \emph{matroid} $\mathcal{M}$ over a finite ground set $\mathcal{P}$ is a collection of independent subset of $\mathcal{P}$. When speaking of a matroid, we denote it as a pair $(\mathcal{P}, \mathcal{M})$, where $\mathcal{M}\subset2^\mathcal{P}$.

\begin{definition}
    We say $\mathcal{M}\subset2^\mathcal{P}$ is a matroid if and only if 
    \begin{enumerate}
    \item $\mathcal{M}\neq \emptyset$; if $\mathcal{R}\in\mathcal{M}$ and $\mathcal{S}\subset\mathcal{R}$ then $\mathcal{S}\in\mathcal{M}$ (downward closed; independence system property)
    \item If $\mathcal{S},\mathcal{R}\in\mathcal{M}$ and $|\mathcal{S}|>|\mathcal{R}|$ then there exists a $p\in\mathcal{S}\backslash\mathcal{R}$ such that $\mathcal{R}\cup\{p\}\in\mathcal{M}$ (augmentation property). 
\end{enumerate}
\end{definition}

Among matroids, the \emph{uniform} and \emph{partition} types stand out for their widespread application.
\begin{definition}[Uniform Matroid]
For some $\kappa\in\mathbb{Z}_{>0}$, the \emph{uniform matroid} over a ground set $\mathcal{P}$ is defined as $(\mathcal{P}, \mathcal{M})$, where $\mathcal{M} = \left\{\mathcal{S}\subset\mathcal{P} \,\big|\, |\mathcal{S}| \leq \kappa\right\}$. 
\end{definition}
A uniform matroid is intuitively understood as allowing any combination of elements, as long as their number does not exceed a specified limit.
\begin{definition}[Partition Matroid]\label{def::partition_matroix}
    For some $\kappa_i\in\mathbb{Z}_{>0}$, $i\in\{1,\cdots,N\}$, the \emph{partition matroid} over a ground set $\mathcal{P}$ is defined as $(\mathcal{P}, \mathcal{M})$, where $\mathcal{M} = \left\{\mathcal{S}\subset\mathcal{P} \,\big|\, |\mathcal{S}\cap\mathcal{P}_i| \leq \kappa_i,~i\in\{1,\cdots,N\}\right\}$, where $\mathcal{P}_1,\cdots,\mathcal{P}_N$ are non-overlapping partition of $\mathcal{P}$, i.e., $\mathcal{P}=\cup_{i=1}^N\mathcal{P}_i$ and $\mathcal{P}_i\cap\mathcal{P}_j=\emptyset$ for any $i\neq j$.
\end{definition}
A partition matroid intuitively groups items into distinct categories, allowing only a specified maximum number of selections from each group.

\section{Application examples}\label{sec::applications}
Many discrete optimal resource allocation problems involve strategic choices to maximize utility, including goals like maximizing entropy, mutual information, symmetric mutual information, cut capacity, weighted coverage, and facility location. As discussed in~\cite{AK-DG:14bookChap}, these utilities have been characterized as submodular functions, permitting the framing of combinatorial optimization involving these utilities as submodular maximization problems. However, adapting some problems to conform to the standard submodular maximization framework often requires pre-processing steps that adjust the decision variables and the ground set to reconfigure the constraints into a matroid format. In what follows, we present some classic examples of submodular maximization problems, which have broad relevance in various decision-making contexts. We also demonstrate through application examples how problem settings can be restructured to align with one of the standard forms of submodular maximization, particularly those subject to uniform or partition matroids.

\medskip
\emph{Exemplar-based clustering}, as introduced by~\cite{LK-PR:09}, aims to identify a subset of exemplars that optimally represent a large dataset by solving the k-medoid problem. This method seeks to minimize the cumulative pairwise dissimilarities between chosen exemplars $\mathcal{P}$ and dataset elements $\mathcal{D}$, formulated as 
\begin{equation}
\label{eq::util_min}
L(\mathcal{R})=\frac{1}{|\mathcal{R}|}\sum\nolimits_{p\in\mathcal{R}}\min_{d\in\mathcal{D}} \text{dist}(p,d),
\end{equation}
for any subset $\mathcal{R}\subset\mathcal{P}$, where $\text{dist}(p,d)\geq 0$, not necessarily symmetric or obeying the triangle inequality, defines the dissimilarity, or distance, between elements. In exemplar clustering, we seek a subset $\mathcal{R}\subset\mathcal{P}$, subject to constraints like cardinality, that minimizes $L$. This problem can be posed as a submodular maximization problem by defining the utility function 
\begin{align}
\label{eq::util_num}
    f(\mathcal{R}) = L(\{d_0\}) - L(\mathcal{R} \cup \{d_0\}),
\end{align}
where $d_0$ is an added hypothetical auxiliary element in the exemplar space. This utility function~\eqref{eq::util_num} quantifies the reduction in the loss associated with the active set versus the loss with merely the phantom placement location, and maximizing this function corresponds to minimizing the loss~\eqref{eq::util_min}. The utility function~\eqref{eq::util_num} is shown to be submodular and monotonically increasing~\cite{RG-AK:10}.

\medskip
Consider now an information harvesting problem where we aim to collect information from a countable set of sources $\mathcal{D}$, distributed across a two-dimensional finite space $\mathcal{W} \subset \mathbb{R}^2$. Suppose we want to deploy $\kappa \in \mathbb{Z}_{>1}$ data harvester devices at a set of pre-specified data retrieval points $\mathcal{B}$ within $\mathcal{W}$, with the condition that $|\mathcal{B}| > \kappa$. We assume the most effective information transfer from an information point $d \in \mathcal{D}$ to a harvester device located at $b \in \mathcal{B}$ occurs when the distance between $b$ and $d$ is minimized. Therefore, for each information point $d \in \mathcal{D}$, the nearest information retrieval point $b \in \mathcal{B}$ with a deployed device is assigned to harvest information. The optimal deployment can be achieved by framing the problem as an exemplar clustering problem with the submodular utility function~\eqref{eq::util_num}, where the distance $\text{dist}(b,d) = \|b - d\|$ represents the Euclidean distance between point $b \in \mathcal{B}$ and data point $d \in \mathcal{D}$. The resulted problem becomes a submodular maximization subject to a uniform matroid, as we aim to select $\kappa$ elements from the ground set $\mathcal{B}$ that maximize utility. In a multi-agent variation of this problem, we envision a scenario where a group of $N$ agents each aims to deploy $\kappa_i\in \mathbb{Z}_{\geq 1}$, $i \in \{1, \cdots, N\}$, data harvesters in $\mathcal{W}$. Every agent $i \in \{1, \cdots, N\}$ has access to a subset $\mathcal{B}_i \subset \mathcal{B}$ of retrieval points, where $\kappa_i < |\mathcal{B}_i|$; the sets of potential retrieval points among agents can overlap. In this setup, the optimal deployment strategy still involves maximizing the submodular utility function~\eqref{eq::util_num}, with $\text{dist}(b,d) = \|b - d\|$. However, the matroid constraint is now a partition matroid, with non-overlapping local ground sets defined for each agent $i\in\{1,\cdots,N\}$ as $\mathcal{P}_i = \{(i, b) \, | \, b \in \mathcal{B}_i\}$. 

\medskip
Next, we shift our focus to another pivotal concept in the landscape of discrete optimization: \emph{the Optimal Welfare problem}~\cite{JV:08}. In the Optimal Welfare problem $m$ items in a discrete set $\mathcal{Q}=\{1,\cdots,m\}$ should be distributed among $N$ agents such that the sum of local utilities of the agents is maximized~\cite{BL-DJL-NN:06}, i.e, 
\begin{align*}\max_{\mathcal{S}_1,\cdots,\mathcal{S}_N\subset\mathcal{Q}} &f(\cup_{i=1}^N\mathcal{S}_i)=\sum\nolimits_{i=1}^Nf_i(\mathcal{S}_i),~\text{s.t.,}\\
&\mathcal{S}_i\cap\mathcal{S}_j=\emptyset, ~~i\neq j,~~i,j\in\{1,\cdots,N\};
\end{align*}
The local cost at each agent, $f_i$, $i\in\{1,\cdots,N\}$ is, normal, monotone increasing and submodular. We know that the sum of submodular functions is submodular, however, optimal Welfare in its original form does not conform to a standard submodular maximization subject to matroid condition. Luckily, this problem can be cast as submodular maximization subject to partition matroid by defining the subsets $\mathcal{P}_i$'s in~\eqref{eq::submdoular_partition} as $\mathcal{P}_i=\{(i,j)\,|\,j\in\{1,\cdots,N\} \}$, $i\in\{1,\cdots,m\}$ and setting $\kappa_i=1$, which limits the assignment of item $i$ to only one agent. In this problem setting, $f^i(\mathcal{R}_i)=f_i(\bar{\mathcal{R}}_i)$ where $\bar{\mathcal{R}}_i=\{j\in\mathcal{Q}\,|\,(i,j)\in\mathcal{R}\}$.

\medskip
Let us now transition to another compelling area of study within set function maximization: sensor placement problems. These scenarios are especially prevalent when operational constraints dictate predetermined locations for deploying sensors. In these problems, we deal with the task of finding the best possible locations for sensors such that best performance is achieved by utility maximization via the limited resources/sensors available. A pertinent instance of this type of problem is detailed in~\cite{NM-RH:18}, focusing specifically on sensor placement within traffic networks in the absence of routing information. The traffic network is represented by a graph $\mathcal{G} = (\mathcal{V},\mathcal{L})$, where $\mathcal{V}$ is the set of network nodes (e.g., intersections), and $\mathcal{L}\subset\mathcal{V}\times\mathcal{V}$ is the set of network links. The traffic network is modeled from a macroscopic point of view, in which for each link $l\in\mathcal{L}$, $f_l$ is the amount of vehicular flow on link $l$ (vehicles per unit of time). The sensor placement objective is to locate sensors, e.g., loop detectors, such that the maximum number of link flows can be identified from a set of measurements at limited number of the nodes. In such a scenario, the goal is to maximize the identifiability of link flows provided that a set of $\kappa$ flow measurements are obtained. Let $\mathcal{L}_m$ be the set of links with flow measurement sensors that deliver measurements $y_l=f_l$ for $l\in\mathcal{L}_m$. Furthermore, we know that at each node, flow conservation must hold. Thus the flow equations consist of $|\mathcal{L}_m|+|\mathcal{V}|$ linear equations, which can be formalized as $A\,f=b$, where $A\in\mathbb{R}^{(|\mathcal{L}_m|+|\mathcal{V}|)\times|\mathcal{L}|}$ is the matrix of linear equations, $f$ is the vector flow across the links and $b$ is the concatenate measurement vectors and the vector $0_{|\mathcal{L}|}$, the right hand value of the conservation of flow equations. Flows considered unidentifiable reside within the null space of matrix $A$. Accordingly,~\cite{NM-RH:18} casts the problem of maximizing the number of identifiable flows as reducing the the null space of A, or, equivalently,
\begin{align}\label{eq::traffic_flow} \max_{\mathcal{L}_m \subset \mathcal{L}} \text{rank}(A), ~~\text{s.t.}~~ |\mathcal{L}_m| \leq \kappa.  
\end{align}
By showing that $\text{rank}(A)$ is a monotone increasing submodular function,~\cite{NM-RH:18} establishes that~\eqref{eq::traffic_flow} is an incidence of the standard submodular maximization subject to uniform matroid. 

\medskip
In some problem settings, it might not be immediately evident that the problem at hand can be framed as set function optimization, especially in terms of submodular maximization subject to matroid constraints. An illustrative example of such a scenario is provided in~\cite{NR-SSK:21}, which addresses a multi-agent persistent monitoring problem over a connected graph through a submodular maximization framework. Specifically,~\cite{NR-SSK:21} examines the persistent monitoring of a set of finite $\mathcal{V}$ interconnected geographical nodes by a finite set of mobile sensors/agents $\mathcal{A}=\{1,\cdots,N\}$, where $|\mathcal{V}|\gg|\mathcal{A}|$. The mobile agents are required to navigate through a set of prespecified edges $\mathcal{E} \subset \mathcal{V} \times \mathcal{V}$, such as aerial or ground corridors, to visit the nodes. The travel time $t_{\mathsf{a},e}$ for each agent $\mathsf{a}\in\mathcal{A}$ across each edge $e\in\mathcal{E}$ is known. Each node $v \in \mathcal{V}$ is assigned a reward function,
\begin{align}\label{eq::Ri}
    R_v(t)=\begin{cases}0, & t=\bar{t}_v,\\
    \psi_v(t-\bar{t}_v), & t>\bar{t}_v,\end{cases}
\end{align}
where $\psi_v(t)$ is a nonnegative, concave, and increasing function of time, and $\bar{t}_v$ represents the most recent time node $v$ has been visited by an agent. In scenarios such as data harvesting or health monitoring, $\psi_v(\cdot)$ might represent the weighted idle time of node $v$, or in event detection, it might indicate the probability of at least one event occurring during the intervals between visits. When any agent $\mathsf{a}\in\mathcal{A}$ arrives at any time $\bar{t}\in\mathbb{R}_{>0}$ at node $v\in\mathcal{V}$, the agent immediately scans the node, and the reward $R_{v}(\bar{t})$ is scored for the patrolling team $\mathcal{A}$; subsequently, $\bar{t}_v$ of node $v$ in~\eqref{eq::Ri} is updated to $\bar{t}$. If more than one agent arrives at node $v\in\mathcal{V}$ and scans it at the same time $\bar{t}$, the reward collected for the team is still $R_v(\bar{t})$. The goal of optimal monitoring involves designing a dispatch policy that determines the tours sequence of interconnected nodes to be visited and the timings of these visits), and the specific agents allocated to each tour to achieve the maximum total reward for the team over the mission horizon $T$. A dispatch policy for an agent $\mathsf{a} \in \mathcal{A}$ over the given mission time horizon is denoted by the tuple $\boldsymbol{\mathsf{p}}_{\mathsf{a}}=(\boldsymbol{\mathsf{V}}_{\mathsf{a}},\boldsymbol{\mathsf{T}}_{\mathsf{a}},\mathsf{a})$, where $\boldsymbol{\mathsf{V}}_{\mathsf{a}}$ and $\boldsymbol{\mathsf{T}}_{\mathsf{a}}$ specify the inter-connected nodes and their corresponding visitation times (tour) assigned to agent $\mathsf{a}$. Given the policies for all agents in the team, $\mathcal{S}=\{\boldsymbol{\mathsf{p}}_1,\cdots,\boldsymbol{\mathsf{p}}_N\}$, the collective reward of the agents is computed as
\begin{align}
\label{eq::reward_global_op}
    \mathsf{R}({\mathcal{S}})=\sum\nolimits_{\forall\, \boldsymbol{\mathsf{p}}\in{\mathcal{S}}}\,\sum\nolimits_{l=1}^{\mathsf{n}_{\mathsf{p}}} R_{\boldsymbol{\mathsf{V}}_{\mathsf{p}}(l)}(\boldsymbol{\mathsf{T}}_{\mathsf{p}}(l)),
\end{align}
where $\mathsf{n}_{\boldsymbol{\mathsf{p}}}$ is the number of nodes to be visited when policy $\boldsymbol{\mathsf{p}}\in\mathcal{S}$ is implemented. The optimal dispatch design, then, is to find $\mathcal{S}$ that maximizes the total reward $\mathsf{R}({\mathcal{S}})$. Notably, overlapping tours, i.e., arrivals at some node(s) $v$ at the same time $\bar{t}$ by more than one agent, are not desired as only one reward $R_v(\bar{t})$ will be collected for the team. For example, for a group of five agents, given any two sets of policies such as $\mathcal{S}_1=\{\boldsymbol{\mathsf{p}}_1,\boldsymbol{\mathsf{p}}_2\}$ and $\mathcal{S}_2=\{\boldsymbol{\mathsf{p}}_3,\boldsymbol{\mathsf{p}}_4,\boldsymbol{\mathsf{p}}_5\}$, we know that  
$$\mathsf{R}(\mathcal{S}_1)+\mathsf{R}(\mathcal{S}_2)\geq \mathsf{R}(\mathcal{S}_1\cup\mathcal{S}_2).$$
Indeed,~\cite{NR-SSK:21} shows that the collective reward function is a submodular set function. It then proceeds to determine the optimal dispatch policy by solving the submodular set function maximization subject to a partition matroid defined by 
\begin{align}\label{eq::persistent_monitoring_submodular_max} \max_{\mathcal{S}\subset\mathcal{P}} \mathsf{R}({\mathcal{S}}), ~~\text{s.t.}~~ |\mathcal{S}\cap\mathcal{P}_{\mathsf{a}}| \leq 1, ~\forall \mathsf{a}\in\mathcal{A}=\{1,\cdots,N\}, 
\end{align}
where $\mathcal{P}_{\mathsf{a}}$ is the set of all feasible policies for agent $\mathsf{a}$, and $\mathcal{P}=\cup_{\mathsf{a}\in\mathcal{A}}\mathcal{P}_{\mathsf{a}}$. When $T$ is large, constructing the entire set of feasible policies for each agent can become prohibitively costly. To manage this cost,~\cite{NR-SSK:21} proposes a moving horizon approach where the problem defined in~\eqref{eq::persistent_monitoring_submodular_max} is considered over a shorter planning horizon. The objective then becomes to identify the optimal policy for this shorter time frame, apply a portion of the policy, and iteratively repeat this process for the remainder of the mission horizon. This strategy enables also a more adaptable and efficient response to the dynamic aspects of the monitoring task.

\section{Solving submodular maximization subject to matroid constraints}\label{sec:central_solutions}
In this section, we give an overview of the two primary approaches to solving submodular maximization subject to matroid constraints: discrete sequential greedy approach and continuous relaxation technique. 

\subsection*{Discrete sequential greedy approach}
Research on problems involving the maximization of submodular 
functions dates back to the work of Nemhauser, Wolsey, and Fisher in the 1970’s~\cite{LMF-GLN-LAW:78,GLN-LAW-MLF:78,GLN-LAW:78}. For monotone-increasing submodular maximization subject to uniform matroid constraint, the celebrated result by Nemhauser et al.~\cite{GLN-LAW-MLF:78} showed that the simple \emph{sequential greedy (SG) algorithm}, which starts at $\mathcal{S}_0=\emptyset$ and iterates according to
\begin{align}
    \mathcal{S}_i=\mathcal{S}_{i-1}\cup \max_{p\in\mathcal{P}\backslash\mathcal{S}_{i-1}}\Delta_f(p|\mathcal{S}_{i-1}),\quad i\in\{1,\cdots,\kappa\},
\end{align}
yields an optimality gap, as defined in~\eqref{eq::optimality_gap}, of $\alpha_{\text{uniform}}=1-\frac{1}{\textup{e}}\approx0.63$, i.e., $f(\mathcal{S}_{\text{SG}})\geq (1-\frac{1}{\textup{e}})\,OPT$, where $\mathcal{S}_{\text{SG}}=\mathcal{S}_\kappa$ and $OPT=f(\mathcal{S}^\star)$. This optimality gap is established elegantly by invoking the monotone increasing and submodularity of the utility function as demonstrated below. In what follows let $\mathcal{S}^\star=\{s_1^\star,\cdots,s_\kappa^\star\}.$
\begin{equation}\label{eq::uniform_bound_proof}
\left.\begin{array}{lll}
    f(\mathcal{S}^\star)&\leq f(\mathcal{S}^\star\cup\mathcal{S}_i)&\text{(monotone increasing)}\\
    &=f(\mathcal{S}_i)+\sum_{j=1}^\kappa \nolimits\Delta_f(s_j^\star|\mathcal{S}_i\cup\{s_1^\star,\cdots,s_{j-1}^\star\})&\text{(telescoping sum)}\\&\leq f(\mathcal{S}_i)+\sum_{j=1}^\kappa \nolimits\Delta_f(s_j^\star|\mathcal{S}_i)&\text{(submodularity)}\\
    &\leq f(\mathcal{S}_i)+\sum\nolimits\nolimits_{j=1}^\kappa(f(\mathcal{S}_{i+1})-f(\mathcal{S}_i)) &\text{(greedy selection to build }\mathcal{S}_{i+1})\\
    &=f(\mathcal{S}_i)+\kappa \,(f(\mathcal{S}_{i+1})-f(\mathcal{S}_i)).&
    \end{array}\right.
\end{equation}
Now defining $\delta_i=f(\mathcal{S}^\star)-f(\mathcal{S}_i)$, we can write $\delta_{i+1}\leq (1-\frac{1}{\kappa})\,\delta_i$. Then, recalling that $\mathcal{S}_0=\emptyset$ and $f(\emptyset)=0$, and using the relationship $(1-\frac{1}{\kappa})^\kappa\leq \frac{1}{\text{e}}$ for any $\kappa\geq 1$, we can show that
\begin{align*}
    f(\mathcal{S}^\star)-f(\mathcal{S}_i)\leq\left(1-\frac{1}{\kappa}\right)^i\left(f(\mathcal{S}^\star)-f(\mathcal{S}_0)\right) \leq \textup{e}^{-i/\kappa}f(\mathcal{S}^\star)\rightarrow  \left(1-\frac{1}{\text{e}}\right)f(\mathcal{S}^\star)\leq f(\mathcal{S}_{\text{SG}}).
\end{align*}
When the total curvature $c$, defined in \eqref{eq::totalCurvature}, of the utility function is known, Conforti and Cornuéjols~\cite{MC-GC:84} refined the characterization of the optimality gap by incorporating this curvature, particularly through the relationship established in \eqref{eq::curvture_inequality2}. From~\eqref{eq::curvture_inequality2}, we can derive
$$f(\mathcal{S}^\star\cup\mathcal{S}_i)-f(\mathcal{S}_i)\geq f(\mathcal{S}^\star)-cf(\mathcal{S}_i).$$
From~\eqref{eq::uniform_bound_proof}, we can express
$$f(\mathcal{S}^\star\cup\mathcal{S}_i)-f(\mathcal{S}_i)\leq  \kappa \,(f(\mathcal{S}_{i+1})-f(\mathcal{S}_i)),$$
Combining these two relationships yields
$$f(\mathcal{S}^\star)-cf(\mathcal{S}_i)\leq  \kappa \,(f(\mathcal{S}_{i+1})-f(\mathcal{S}_i)),$$
which can be rewritten as
$$f(\mathcal{S}^\star)-cf(\mathcal{S}_i)\leq \frac{\kappa}{c} \,\left(cf(\mathcal{S}_{i+1})-cf(\mathcal{S}_i)\right).$$
Defining $\delta_i=f(\mathcal{S}^\star)-cf(\mathcal{S}_i)$, we can express
$\delta_{i+1}\leq \left(1-\frac{c}{\kappa}\right)\,\delta_i.$
Recalling that $\mathcal{S}_0=\emptyset$ and $f(\emptyset)=0$, we can demonstrate that
\begin{align*}
    f(\mathcal{S}^\star)-cf(\mathcal{S}_i)&\leq\left(1-\frac{c}{\kappa}\right)^i\,\left( f(\mathcal{S}^\star)-cf(\mathcal{S}_0)\right) \leq \textup{e}^{-i\,c/\kappa}f(\mathcal{S}^\star) \rightarrow  \frac{1}{c}\left(1-\frac{1}{\text{e}^c}\right)f(\mathcal{S}^\star) \leq f(\mathcal{S}_{\text{SG}}).
\end{align*}
These derivations result in the optimality gap for uniform matroids of  $\alpha_{\text{uniform}}=\frac{1}{c}\left(1-\frac{1}{e^c}\right).$
\medskip

The curvature $c$ quantifies the degree of diminishing returns exhibited by a set function. At one extreme, $c=0$ corresponds to a modular function, where $f(\{p_1,p_2\})= f(\{p_1\})+f(\{p_2\})$ for any $p_1,p_2\in\mathcal{P}$. In this case, $\alpha_{\text{uniform}}=1$, indicating that the sequential greedy algorithm achieves the optimal solution within a finite number of steps for modular functions under any matroid constraint. At the other extreme, $c=1$ implies the existence of at least one element that, under certain circumstances, provides no marginal benefit to the function $f$. This represents the most challenging scenario for the greedy algorithm. In practical applications where the total curvature cannot be determined precisely, it is prudent to adopt a conservative approach by assuming $c=1$. This ensures that the performance guarantees hold even in the most demanding situations, albeit at the cost of potentially underestimating the algorithm's effectiveness in more favorable cases.

\medskip
Owing to the intrinsic matroid characteristics such as downward-closeness and augmentation, the sequential greedy algorithm, initially developed for problems constrained by uniform matroids, can indeed be adapted to tackle submodular maximization under any matroid constraint. These fundamental properties ensure that the algorithm maintains the feasibility of the solution at each step and guarantees that the provided solution is both suboptimal and feasible, thereby highlighting the broad applicability and efficacy of the greedy method across a multitude of optimization scenarios. It has been demonstrated that, for monotone increasing submodular utility functions within an optimization problem~\eqref{eq::gen_set_opt} where $\mathcal{F}(\mathcal{P})$ is a matroid, i.e., $\mathcal{F}(\mathcal{P})=(\mathcal{P}, \mathcal{M})$, the sequential greedy algorithm 
\begin{align}
\mathcal{S}_{i} = \mathcal{S}_{i-1} \cup \!\!\! \max_{p \notin \mathcal{S}_{\text{SG}}, \mathcal{S}_{i-1} \cup \{p\} \in \mathcal{M}} \!\!\!\!\!\!\! \Delta_f(p | \mathcal{S}_{i-1}),
\end{align}
proceeds until no additional $p$ makes $\mathcal{S}_{i-1} \cup \{p\}$ a feasible selection within $\mathcal{M}$. This process is guaranteed to yield a solution with an optimality gap of $\alpha = \frac{1}{2}$, that is, $f(\mathcal{S}_{\text{SG}}) \geq \frac{1}{2} \cdot OPT$, where $\mathcal{S}_{\text{SG}} = \mathcal{S}_\kappa$ and $OPT = f(\mathcal{S}^\star)$. Specifically, in the context of submodular maximization subject to a partition matroid~\eqref{eq::submdoular_partition}, the guaranteed optimality gap achieved by the sequential greedy algorithm is documented as $\alpha_{\text{partition}} = \frac{1}{2}$.

\medskip
For submodular maximization subject to a partition matroid~\eqref{eq::submdoular_partition}, the sequential greedy algorithm can be implemented more efficiently by making sequential greedy choices starting from $\mathcal{P}_1$, choosing $\kappa_1$ elements from this set, then fixing these choices before proceeding to set $\mathcal{P}_2$ to choose the $\kappa_2$ elements, and so on, until the last $\kappa_N$ choices are made after fixing the first $\kappa_1, \ldots, \kappa_{n-1}$ choices. The process can be explained as  starting with $\mathcal{S}_0=\emptyset$ and $\mathcal{R}_0=\emptyset$ and iterating over 
\begin{align}\label{eq::sequential_greedy_partition_matroid}
\mathcal{S}_{i}=\, &\mathcal{S}_{i-1}\cup \mathcal{R}_{\kappa_i},\quad i\in\{1,\cdots,N\}, \\
   & \mathcal{R}_j=\mathcal{R}_{j-1}\cup \max_{p\in\mathcal{P}_i\backslash\mathcal{R}_{i-1}}\Delta_f(p|\mathcal{S}_{i-1}\cup\mathcal{R}_{j-1}),\quad j\in\{1,\cdots,\kappa_i\}; \nonumber
\end{align}
this process outputs $\mathcal{S}_{\text{SG}}=\mathcal{S}_N$. 
To illustrate the role of submodularity in setting the optimality gap of $\alpha_{\text{partition}} = \frac{1}{2}$ for the sequential greedy algorithm~\eqref{eq::sequential_greedy_partition_matroid}, next, we provide a concise overview of the proof. For simplicity, let us assume $\kappa_i = 1$, indicating our goal to select one element from each subset $\mathcal{P}_i$, for all $i \in \{1, \cdots, N\}$ in~\eqref{eq::submdoular_partition}. Mirroring the steps from the proof of the optimality gap in submodular maximization under a uniform matroid, we can establish, given that the utility function is normal, monotone increasing, and submodular, the following inequality:
\begin{equation}\label{eq::proof_partition_matroid}
    f(\mathcal{S}^\star)\leq  f(\mathcal{S}_{N})+\sum\nolimits_{i=1}^N \Delta_f(s_i^\star|\mathcal{S}_N),
\end{equation}
where $s_1^\star,\cdots,s_N^\star$ are the elements of $\mathcal{S}^\star$, that is, $\mathcal{S}^\star=\{s_1^\star,\cdots,s_N^\star\}$. Let $\bar{s}_1,\cdots,\bar{s}_N$ be the elements of $\mathcal{S}_N$, that is $\mathcal{S}_N=\{\bar{s}_1,\cdots,\bar{s}_N\}$. Note that $s_i^\star$ and $\bar{s}_i$ belong to $\mathcal{P}_i$ for each $i\in\{1,\cdots,N\}$. Thus, we can write
\begin{equation*}
\left.\begin{array}{lll} \Delta_f(s_{i}^\star|\mathcal{S}_N)&=f(\mathcal{S}_{i-1}\cup\{\bar{s}_{i},\cdots,\bar{s}_N\}\cup \{s_{i}^\star\})-f(\mathcal{S}_{i-1}\cup\{\bar{s}_{i},\cdots,\bar{s}_N\})
    \\&\leq f(\mathcal{S}_{i-1}\cup \{s_i^\star\})-f(\mathcal{S}_{i-1})&\!\!\!\!\!\!\!\!\!\!\!\!\!\!\!\!\text{(by submodularity)}\\
    &\leq f(\mathcal{S}_{i})-f(\mathcal{S}_{i-1}) &\!\!\!\!\!\!\!\!\!\!\!\!\!\!\!\!\text{(by greedy selection to construct~}\mathcal{S}_i\text{)}.
    \end{array}\right.
\end{equation*}
Therefore, from~\eqref{eq::proof_partition_matroid}, we conclude that $f(\mathcal{S}^\star)\leq 2 f(\mathcal{S}_{\text{SG}})$, thus establishing that $\alpha_{\text{partition}} = \frac{1}{2}$. It is important to note that this optimality gap represents the worst-case scenario and is independent of the order in which the disjoint components $\mathcal{P}_1,\cdots,\mathcal{P}_N$ of the ground set $\mathcal{P}$ are engaged in the sequential greedy selection process outlined in~\eqref{eq::sequential_greedy_partition_matroid}.

When the total curvature $c$, defined in \eqref{eq::totalCurvature}, of the utility function is known, we can refine our characterization of the optimality gap for the partition matroid constraint, as well. Conforti and Cornuéjols~\cite{MC-GC:84} established that for submodular maximization under a partition matroid constraint, the optimality gap is $\alpha_{\text{partition}} = \frac{1}{1+c}$. This result provides a tighter bound than the general $1/2$ approximation when $0\leq c < 1$. As discussed earlier, $c=0$ corresponds to a modular function, resulting in $\alpha_{\text{partition}}=1$, indicating that the sequential greedy algorithm achieves the optimal solution. Conversely, when $c=1$, we recover the standard $1/2$ approximation ratio.

\subsection*{Continuous relaxation: the continuous sequential greedy method}
While the discrete sequential greedy algorithm provides an efficient and practical solution to submodular maximization problems subject to general matroid constraints, it is limited by its optimality gap guarantee of $\alpha = \frac{1}{2}$. To push beyond these limits, researchers have developed more sophisticated approaches that leverage continuous optimization techniques. One of the most powerful of these is the multilinear extension method, which transforms the discrete optimization problem into a continuous one, allowing for the application of advanced optimization techniques.

The literature has shown that for monotone increasing submodular function maximization subject to a matroid constraint, it is computationally hard to approximate this problem within a factor better than $1-1/e\approx 0.63\%$~\cite{UF:98}. A suboptimal solution, aiming at achieving such bound for submodular maximization subject to partition matroid is proposed in~\cite{JV:08} and explained in further detail in~\cite{GC-CC-MP-JV:11}. The approach introduced by~\cite{JV:08} utilizes the multilinear extension of submodular set functions, as defined in~\eqref{eq::F_determin}, along with a continuous relaxation of the partition matroid, to solve for a fractional solution of the continuous relaxation of~\eqref{eq::submdoular_partition}. A suboptimal solution for~\eqref{eq::submdoular_partition} is then recovered through a lossless rounding procedure. In the following, we provide a brief overview of this sophisticated approach.

\medskip
To simplify our notation, in the remainder of this article, we assume without loss of generality that the ground set is given by $\mathcal{P}=\bigcup\nolimits_{j=1}^N  \mathcal{P}_j=\{1,\cdots,n\}$ and for any $i\in\mathcal{A}$ we have $\mathcal{P}_i=\{l,l+1,\cdots,m
\}\subset\{1,\cdots,n\}$ and $\mathcal{P}_{i+1}=\{m+1,m+2,\cdots,q
\}\subset\{1,\cdots,n\}$.

\medskip

The \emph{matroid polytope}, which is the continuous extension of the partition matroid described in Definition~\ref{def::partition_matroix}, is 
\begin{align}\label{eq::convexhull}P(\mathcal{M})=\left\{ \boldsymbol{x} \in [0,1]^{n}\,\big|\,\sum\nolimits_{p \in \mathcal{P}_i}{}[\boldsymbol{x}_i]_{p}\leq \kappa_i, \forall i\in\mathcal{A}\right\}.
\end{align}
Let $\boldsymbol{x}=[\boldsymbol{x}_1^\top,...,\boldsymbol{x}_N^\top]^\top$, where $\boldsymbol{x}_i \in [0,1]^{|\mathcal{P}_i|}$ is the membership probability vector of $\mathcal{P}_i$ that defines the probability of choosing policies from $\mathcal{P}_i$ for each agent $i\in\mathcal{A}$. Then, the matroid polytope~\eqref{eq::convexhull} is the convex hull of the partition matroid in the space of the membership probability vector. 

\medskip
With the multilinear extension and matroid polytope defined, the continuous relaxation of problem~\eqref{eq::submdoular_partition} is expressed as 
 \begin{align}\label{eq::conti-approx}
\max F(\boldsymbol{x}),~~\textup{s.t.,}~~\boldsymbol{x}\in P
(\mathcal{M}).
\end{align} 
\cite{JV:08} has shown that $OPT=\max_{\boldsymbol{x}\in P(\mathcal{M})} F(\boldsymbol{x})$, $OPT$ is the optimal value of the submodular maximization problem~\eqref{eq::submdoular_partition}. Therefore, by finding the maximizer of~\eqref{eq::conti-approx} and employing a lossless rounding, we can determine the submodular maximization problem~\eqref{eq::submdoular_partition}. But, optimization problem~\eqref{eq::conti-approx} is non-convex/non-concave  maximizer, which implies that numerical solutions to this problem do not always guarantee convergence to the global maximizer. 

\medskip
The solution proposed by~\cite{JV:08} for~\eqref{eq::conti-approx}, known as the *continuous greedy* algorithm, involves moving in the direction of maximum ascent within $P(\mathcal{M})$ by following the flow:
\begin{align}\label{eq::VNabF}
\frac{\text{d}\boldsymbol{x}}{\text{d}t}=\boldsymbol{v}(\boldsymbol{x})~~\text{where}~~ \boldsymbol{v}(\boldsymbol{x})=\underset{\boldsymbol{w}\in P(\mathcal{M})}{\arg\max}(\boldsymbol{w}\cdot\nabla F(\boldsymbol{x})),
\end{align} 
over the time interval $[0,1]$. This approach can be seen as a variation of the so-called Frank-Wolfe algorithm, also referred to as the conditional gradient, introduced by~\cite{frank1956algorithm}. The Frank-Wolfe algorithm is a straightforward first-order iterative constrained optimization algorithm designed for optimizing smooth functions over closed, bounded convex sets.

\medskip
\cite{JV:08} showed that $\boldsymbol{x}(t)$, generated by~\eqref{eq::VNabF}, stays within $P(\mathcal{M})$ for $t \in [0,1]$, as it forms a convex combination of vectors in $P(\mathcal{M})$. Additionally, $\boldsymbol{x}(1)$ reaches the boundary of $P(\mathcal{M})$. \cite{JV:08} further showed that by following~\eqref{eq::VNabF}, we achieve $F(\boldsymbol{x}(1)) \geq (1-1/\text{e})\,F(\boldsymbol{x}^\star)$, and thus $F(\boldsymbol{x}(1)) \geq (1-1/\text{e})\,OPT$, where $OPT=f(\mathcal{S}^\star)$. Next, by employing the Pipage rounding method proposed by~\cite{AAA-MIS:04} or its stochastic variant as described in, the fractional solution $\boldsymbol{x}(1)$ is transformed into an integral point $\bar{\boldsymbol{x}}$ within $(\mathcal{P},\mathcal{M})$. This integral solution $\bar{\boldsymbol{x}}$, represented as $[\bar{\boldsymbol{x}}_1^\top,\cdots,\bar{\boldsymbol{x}}_N^\top]^\top$ and confined to the set $\{0,1\}^n$, meets the condition $F(\bar{\boldsymbol{x}}) \geq F(\boldsymbol{x}(1))$. As a result, $\bar{\mathcal{S}} = \mathcal{S}_{\bar{\boldsymbol{x}}}$ within $(\mathcal{P},\mathcal{M})$ constitutes a deterministic feasible set that not only satisfies $f(\bar{\mathcal{S}}) \geq F(\boldsymbol{x}(1))$ but also ensures $f(\bar{\mathcal{S}}) \geq (1-1/\text{e})OPT$. While the specifics of the Pipage rounding method are omitted here for brevity, it is crucial to acknowledge that the ability to reach the boundary of $P(\mathcal{M})$—that is, 
\begin{align*}
\boldsymbol{x}(1)\in\left\{ \boldsymbol{x} \in [0,1]^{n}\,\big|\,\sum\nolimits_{p \in \mathcal{P}_i}{}[\boldsymbol{x}_i]_{p}= \kappa_i, \forall i\in\mathcal{A}\right\},
\end{align*}
plays a pivotal role in achieving lossless rounding through the Pipage rounding method.

\medskip

Although elegant, a number of issues need to be addressed regarding the continuous greedy algorithm~\eqref{eq::VNabF} to make it a practical and computationally reasonable solution. One challenge is computing the conditional gradient $\boldsymbol{v}(\boldsymbol{x})$ in~\eqref{eq::VNabF}, which, at first glance, seems to require solving a linear program. However, given that the gradient $\nabla F(\boldsymbol{x})$ consists of nonnegative elements, one solution for $\boldsymbol{v}(\boldsymbol{x}) = [\boldsymbol{v}_1^\top,...,\boldsymbol{v}_N^\top]^\top$ is $\boldsymbol{v}_i \in \{0,1\}^{|\mathcal{P}_i|}$ with $\boldsymbol{v}_i=\boldsymbol{1}_{\boldsymbol{C}_i}$, where $\boldsymbol{C}_i$ is the set of the indexes of $\kappa_i$ largest elements of $\nabla F(\boldsymbol{x})_i$. Here, we partitioned $\nabla F(\boldsymbol{x})$ as $[\nabla F(\boldsymbol{x})_1^\top,...,\nabla F(\boldsymbol{x})_N^\top]^\top$.

\medskip
The next challenge addressed is adopting a more practical solution through the use of a numerical iterative process:
\begin{align}\label{eq::xv}
    \boldsymbol{x}(t+1)=\boldsymbol{x}(t)+\frac{1}{T}\boldsymbol{v}(t), ~~\text{where}~~ \boldsymbol{v}(t)=\underset{\boldsymbol{w}\in P(\mathcal{M})}{\arg\max}(\boldsymbol{w}\cdot\nabla F(\boldsymbol{x})),
\end{align}
where $1/T$, with $T\in\mathbb{Z}_{>0}$, is the step size used to discretize $[0,1]$ instead of following the continuous path in~\eqref{eq::VNabF}. However, a more significant issue is constructing the gradient $\nabla F(\boldsymbol{x})$ without the need to compute $f(\mathcal{R})$ for all $\mathcal{R} \in 2^{\mathcal{P}}$. Given that multilinear extension requires this information, computing $\nabla F(\boldsymbol{x})$ becomes computationally intractable as the size of the ground set $\mathcal{P}$ increases. A practical solution is provided by considering the stochastic interpretation~\eqref{eq::firstDer} of the gradient. Drawing enough set samples according to the membership probability vector $\boldsymbol{x}$ can yield an estimate of $\nabla F(\boldsymbol{x})$ at a reasonable computational cost. The Chernoff-Hoeffding inequality~\cite{WH:94} can be utilized to determine the quality of this estimation given the number of samples.~\cite{JV:08} shows that the optimality gap for this practical approach is $1-1/\text{e}-O(1/T)$ with the probability of $1-2Tn\text{e}^{-\frac{1}{8T^2}{K}}$, where $K$ is the number of samples. Further studies on the use of the multilinear extension for submodular maximization are discussed in~\cite{JV:10,GC-CC-MP-JV:11,CC-JV-RC:14,AAB-BM-JB-AK:17,AM-HH-AK:20,OS-MF:20}.

\section{Distributed Submodular maximization}\label{sec::distributed}
In many submodular maximization problem settings, the volume of collected data is not only vast but also expanding rapidly. The widespread deployment of sensors, for example, has resulted in the collection of large quantities of physical world measurements. Similarly, data on mechanical and human activities are being captured and stored at ever-increasing rates and with greater detail. These datasets are often complex and high-dimensional, necessitating distributed storage and processing solutions. Efforts to adapt the sequential greedy algorithm for tackling large-scale submodular maximization problems have included attempts to reduce the problem size through approximations~\cite{KW-RI-JB:14}. Given the extensive algorithmic challenges, parallel computing naturally lends itself as a solution, with certain distributed algorithmic solutions for submodular optimization already explored, primarily focusing on submodular maximization subjected to uniform matroid constraints; see e.g.,~\cite{BM-AK-RS-AK:13,RB-AE-HLN-JW:15,BM-AK-RS-AK:16,BM-MZ-AK:16,RK-BM-SV-AV:15,PSR-OS-MF:20}. Many of these efforts utilize the MapReduce method~\cite{JD-SG:08}, which is arguably one of the most successful models for reliable and efficient parallel computing. Within this method, the ground set is distributed among $m$ independent machines (the map phase), each with limited memory and processing power. These machines then perform computations in parallel on their allotted data subsets. The outcomes of these computations are then collated onto a single machine and further processed in the final round to produce the result. It is during the shuffle phase that these machines can communicate and exchange data.

\medskip
While parallel computing plays a crucial role in addressing the challenges of large-scale data, the shift towards distributed implementations of submodular maximization extends beyond the pursuit of computational efficiency. In some submodular maximization problems especially in cyber-physical system (CPS) and networked system  applications, the elements of the optimization problem—such as the elements of the utility function or the discrete decision set—are often distributed across multiple entities (agents), which, despite seeking the optimal solution, might not want to share their information with a central authority. For such applications, the solution is desired in a distributed manner, where agents arrive at the solution using local interactions. Many challenges must be faced when designing distributed submodular maximization solvers. The algorithmic challenges stem from processing the data that is distributed across several machines/agents while using minimal communication and synchronization across the agents/machines. At the same time, it is important to deliver solutions that are competitive with the centralized solution when all fragmented information is available to a single authority.

\begin{figure}[t]
\begin{center}
    {\includegraphics[width=.7\textwidth]{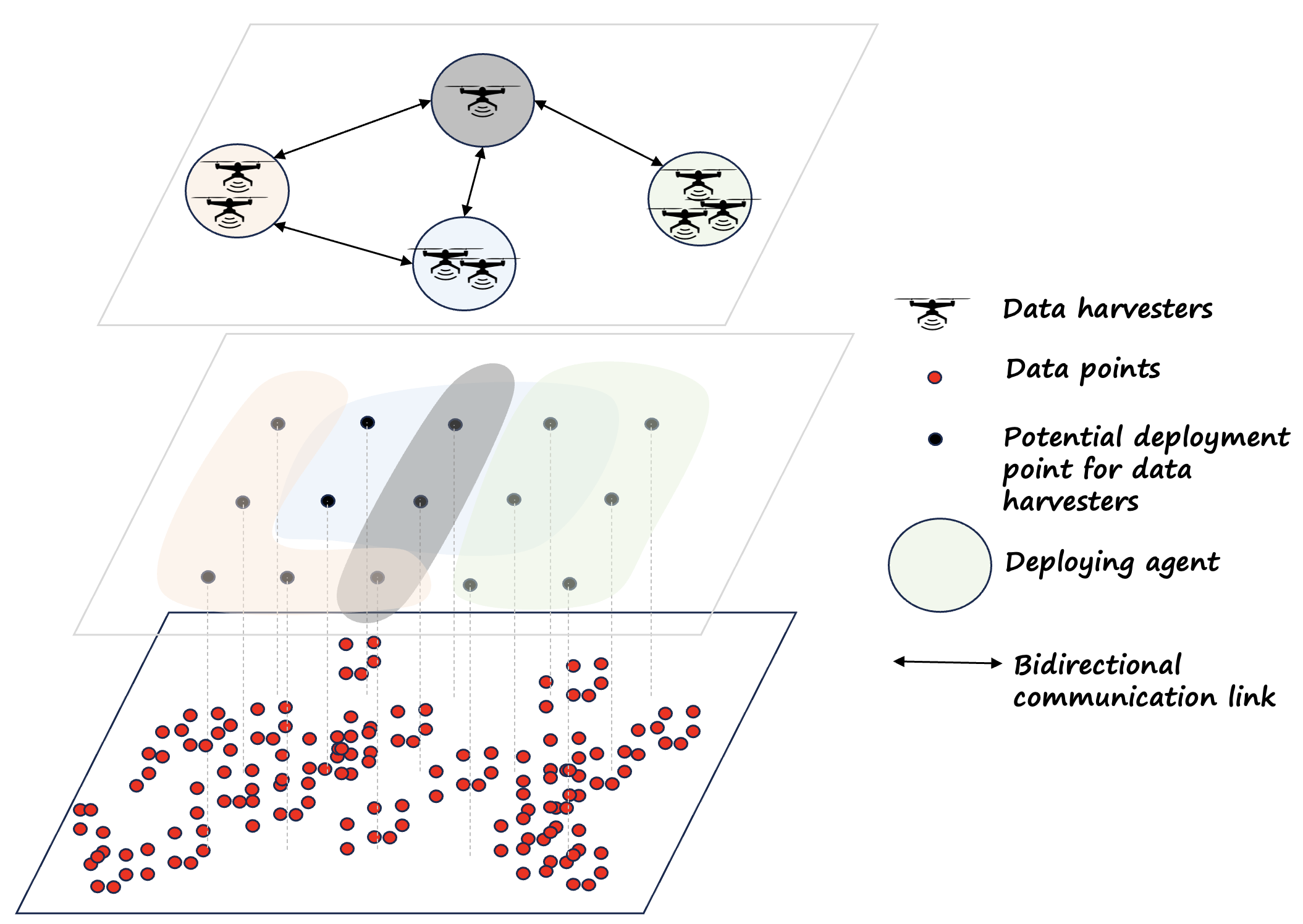}}%
    \caption{{\small   
A simple schematic illustration of a distributed sensor deployment for data harvesting involving a group of agents $\mathcal{A}$. Each agent $i \in \mathcal{A}$ possesses $\kappa_i$ data harvesting drones, which can be deployed at a set of pre-assigned deployment points $\mathcal{B}_i$, indicated by the same color used to depict the agent.  The agents communicate over a connected graph to determine the optimal deployment positions for the team, as defined by the submodular maximization problem subject to a partition matroid described in Section~\ref{sec::applications}.
}}\label{fig::data_harvest_distributed}
\end{center}
\end{figure}

\medskip
Due to space limitations, we will elaborate only on the solution approaches for a class of distributed submodular maximization classified as \emph{distributed ground set}, which is an instance of submodular maximization subject to partition matroid. In this class of problems, the disjoint sets $\mathcal{P}_1, \cdots, \mathcal{P}_N$ each respectively belong to an agent $i \in \mathcal{A} = \{1, \cdots, N\}$ and are \emph{not known} to other agents. The number of strategy choices of each agent $i$, $\kappa_i$, is also known \emph{only} to the respective agent. The agents' access to the utility function is through a black box that returns $f(\mathcal{R})$ for any given set $\mathcal{R} \in \mathcal{P}$ (value oracle model). As an example, consider the data harvesting problem discussed in Section~\ref{sec::applications}, where we have a group of agents $\mathcal{A}$, which communicate over a connected undirected graph and want to decide via local interactions where to deploy their data harvester devices to maximize the utility function~\eqref{eq::util_num}; see Fig.~\ref{fig::data_harvest_distributed}.

\begin{figure}[t]
\begin{center}    {\includegraphics[width=.7\textwidth]{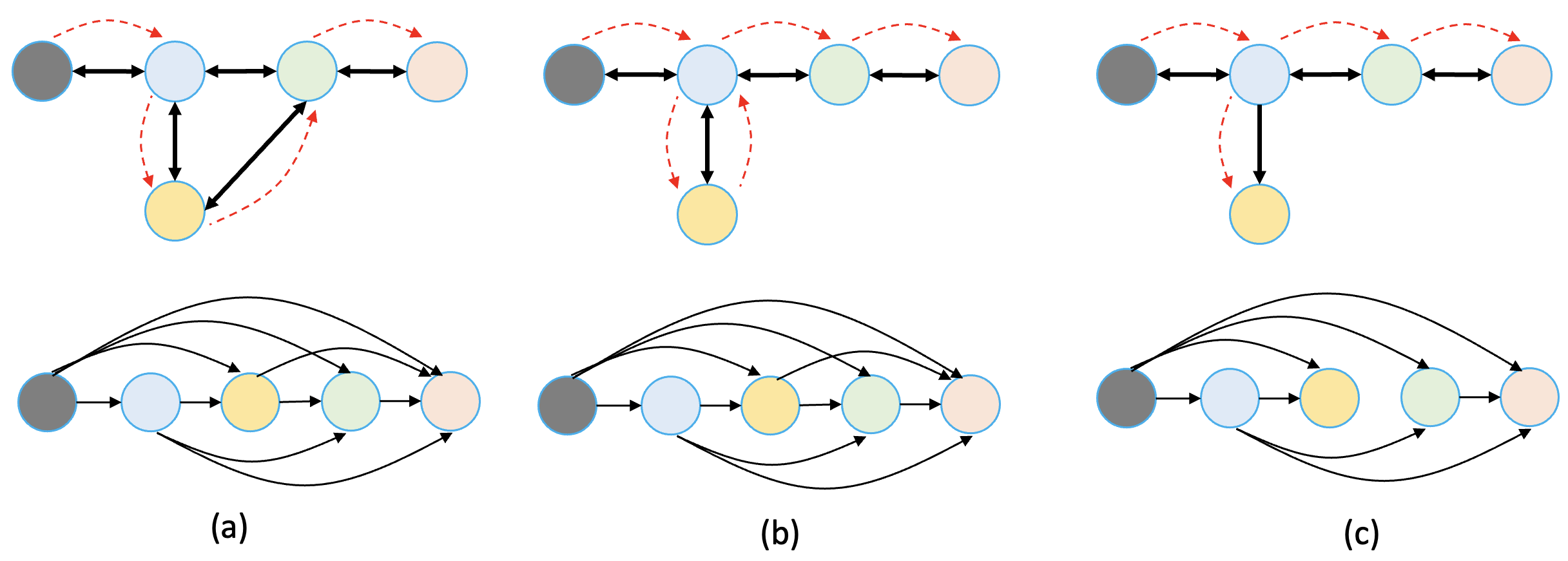}}%
    \caption{{\small   
The top figures illustrate the communication topology (solid black arrows) and the message-passing sequence (dashed red arrows). The bottom figures detail the information-sharing graph. In case (a), the communication graph possesses a Hamiltonian path, enabling the shortest sequence for optimal message passing in the sequential greedy algorithm. In case (b), the absence of a Hamiltonian path in the communication graph requires the blue agent to participate twice in the message-passing process to implement the sequential greedy algorithm. In case (c), with the communication graph disconnected, the information-sharing graph becomes incomplete, preventing the precise execution of the sequential greedy algorithm. Consequently, the guaranteed optimality gap decreases from $1/2$.
}}\label{fig::massage_pass}
\end{center}
\end{figure}

\medskip
For distributed ground set problems, the sequential greedy algorithm, described in~\eqref{eq::sequential_greedy_partition_matroid}, readily adapts to decentralization, either through sequential message-passing or by sharing messages sequentially via the cloud, see~\cite{NR-SSK:21}. However, decentralizing the sequential greedy algorithm introduces communication routing overhead. When agents communicate over a connected graph, implementing sequential message-passing necessitates identifying a Hamiltonian path—a path that visits each agent on the graph exactly once—which is an NP-hard problem. If a Hamiltonian path does not exist within a graph, an alternative path that visits agents the fewest number of times must be identified to ensure communication-efficient sequential message-passing. This concept is visually represented in Figure~\ref{fig::massage_pass}(a) and (b), demonstrating scenarios with and without a Hamiltonian path, respectively. Moreover, it has been demonstrated that the sequence order affects the actual approximation factor of the solution derived from the sequential greedy algorithm~\cite{RK-DG-JM:21}; see also~\cite{NR-SSK:23} for further numerical examples. The difficulty of pinpointing the sequence that yields the best solution grows exponentially with the size and connectivity of the communication network. 

\medskip
In the implementation of the sequential greedy algorithm through message-passing, an additional key factor to consider is the impact of message drop-offs, or incomplete message-passing sequences, on the optimality gap. Such drop-offs lead to an incomplete information sharing graph, $\mathcal{G}_I$. The lower plots in Fig.~\ref{fig::massage_pass} illustrate this graph, visualizing the flow of information as a result of the message-passing process. Here, an arrow from agent $i$ to agent $j$ signifies successful information relay from $i$ to $j$, whether directly or indirectly through preceding agents. Research by~\cite{BS-SLS:18} reveals that the optimality gap is intricately linked to the clique number, $\mathcal{W}(\mathcal{G}_I)$, of the information graph. Specifically, the clique number represents the size of the largest fully connected component in the information graph's undirected version. The findings demonstrate that the optimality gap expands as communication paths become disjointed, a phenomenon quantified as $
    f(\bar{\mathcal{S}}_{SG}) \geq \frac{1}{2+n-\mathcal{W}(\mathcal{G}_I)} f(\mathcal{S}^\star)$.
Notably, this analysis guarantees reaching the well-known optimality gap of $1/2$ for the problem of submodular maximization under partition matroid constraints when the information graph is complete and $\mathcal{W}(\mathcal{G}_I)=n$, highlighting the critical role of information graph connectivity in algorithm performance.

\medskip
Extending this line of research,~\cite{JV-SSK:24} introduces a probabilistic message-passing framework for decentralized submodular maximization. This approach considers the probability of successful communication between agents in each step of the sequential process, rather than assuming deterministic message delivery. The study introduces the concept of a probabilistic optimality gap, which provides insights into designing effective message-passing sequences and determining which agents should be given additional communication opportunities in resource-constrained environments. This probabilistic perspective offers a more realistic model for analyzing algorithm performance in uncertain communication settings.

\medskip
The distributed implementation of the continuous greedy algorithm~\eqref{eq::xv} over connected graphs has been explored in the literature. However, for the class of problems where the ground set is distributed, results are relatively scarce. A notable contribution in this area was proposed by \cite{AR-AA-BS-JGP-HH:19} and \cite{NR-SSK:23}. Specifically, for the special class of submodular set functions with curvature \(c=1\), and when each agent is restricted to choosing only a single strategy from its own strategy set (\(i.e., \kappa_i=1\)), \cite{AR-AA-BS-JGP-HH:19} introduced an average consensus-based distributed algorithm to address the maximization problem over connected graphs. The approach requires a closed-form expression of the multilinear extension function. However, constructing the closed form of the multilinear extension of a submodular function and its derivatives exponentially increases in computational complexity relative to the size of the strategy set. Moreover, this result also depends on a centralized rounding scheme. An alternative solution employing a max-consensus method for local interaction among agents was proposed in \cite{NR-SSK:23}. This method takes advantage of a sampling-based approach to approximate $\nabla F(\boldsymbol{x})$ and employs a distributed stochastic rounding procedure, allowing agents to locally derive their policy selection, thus achieving a practical and fully distributed solution. We omit the technical details of these distributed solutions for brevity. Nevertheless, it is essential to highlight a significant hidden challenge in the distributed continuous greedy algorithms that utilize the multilinear extension of the utility function. In the distributed implementations of the continuous greedy algorithm~\eqref{eq::xv} as seen in~\cite{AR-AA-BS-JGP-HH:19} and~\cite{NR-SSK:23}, each agent maintains a local copy of the membership probability vector $\boldsymbol{x}$ and employs local interactions through consensus protocols to align these local copies towards a common choice that approximates the central solution. However, an inherent challenge involves the local computation of $\nabla F(.)$, which necessitates access to all elements of any other agent’s local ground set for which the corresponding element of the local copy of $\boldsymbol{x}$ is non-zero. This necessity implies that the distributed implementation of the continuous greedy algorithm involves not only the local exchange of $\boldsymbol{x}$ among neighboring agents but also the sharing of elements of the local ground sets of the agents across the network. \cite{NR-SSK:23} proposes a mechanism for managing this information sharing. For further details, interested readers are encouraged to refer to \cite{NR-SSK:23}.

\section{Concluding Remarks}\label{se:conclusion}
In conclusion, this article has systematically explored submodular maximization problems subject to uniform and partition matroids, uncovering their significant applications across various disciplines. By diving into algorithmic solutions like the sequential greedy algorithm and its adaptations for distributed contexts, we highlight the vast potential for key advancements in optimization theory and practical problem-solving. Further attention to the continuous greedy algorithm opens a gateway to addressing the computational intricacies inherent in these problems, offering a refined tool for approaching submodular maximization within a continuous setting. This combined exploration of discrete and continuous methods enriches the field's theoretical base while expanding its applicability to real-world challenges, affirming the pivotal role of matroid constraints in shaping solution strategies for submodular maximization. 

\medskip
Submodular maximization theory, particularly when subjected to matroid constraints, is an expanding field that continues to attract considerable interest.  We close this article with a few recommendations for further reading for the interested reader.

\subsection{Further reading}
Submodularity, a defining characteristic of set functions, provides a theoretical framework that aids in developing systematic approaches to tackle complex optimization challenges. However, not all utility functions inherently display submodular characteristics. This observation has prompted researchers to extend their inquiries beyond classical submodularity, exploring variations such as weakly or proportional submodularity as introduced in~\cite{das2011submodular}, along with lattice submodularity detailed in~\cite{soma2018maximizing}. These lines of investigation have significantly broadened the theoretical landscape, enabling the application of optimization algorithms to a more diverse array of problems where traditional submodularity is not directly applicable. For instance, applications leveraging these variants are discussed in~\cite{hashemi2020randomized} and \cite{liu2018controlled}.

\medskip
Similarly, the concept of $k$-submodularity extends the principle of submodularity to scenarios where decision-making involves multiple states beyond simple binary choices. Foundational contributions to this topic, such as those by \cite{huber2012towards} and~\cite{ward2016maximizing}, illustrate how $k$-submodularity broadens the traditional approach by accounting for functions in which elements can occupy one of $k+1$ states, not limited to mere presence or absence. This broader perspective supports more nuanced problem modeling, especially in situations where elements engage at varying levels or capacities.

\medskip
Furthermore, while sequential greedy and sampling-based approximate continuous greedy algorithms offer polynomial-time solutions, their computational demands can escalate with larger ground sets. Here, the ``lazy" evaluation strategy, which aims to minimize computational load by circumventing unnecessary function evaluations, plays a critical role in enhancing efficiency. As initially proposed by~\cite{MM:78} and further refined in the ``Lazier than Lazy Greedy" by~\cite{mirzasoleiman2015lazier}, these approaches significantly streamline the process of maximizing submodular functions. They cleverly leverage the diminishing returns property, updating marginal gains only for top-priority elements, thereby curbing the requisite number of function evaluations.

\medskip
Besides computational efficiency, modern applications of submodular maximization, particularly in distributed environments, require privacy preservation. For instance, in distributed operations, agents need assurance that their policy choices will remain confidential. While distributed approaches reduce central data aggregation, communications between agents could still expose distributed network operations to adversarial eavesdroppers. Research into privacy for submodular maximization employs methods ranging from differential privacy as in~\cite{fouad2014supermodularity} to novel strategies like the one by~\cite{rezazadeh2022distributed}, which leverages the stochastic rounding in continuous greedy algorithms for privacy guarantees.

\medskip
The field of submodular maximization is rapidly advancing, continually extending the boundaries of optimization theory. Key emerging topics that exemplify the forefront of research in this domain include Deep Submodular functions~\cite{dolhansky2016deep}, Online submodular maximization~\cite{xu2023online}, Mixed-integer programming approaches to generalized submodular optimization~\cite{kuccukyavuz2023mixed}, Adaptive Submodular maximization~\cite{esfandiari2021adaptivity}, Streaming submodular maximization~\cite{badanidiyuru2014streaming}, Submodular reinforcement learning~\cite{prajapat2023submodular}, and Fairness in submodular maximization~\cite{el2024fairness}.

\end{document}